\documentclass[preprint,prd,amsmath,amssymb,amsthm,nofootinbib]{revtex4}
\usepackage[mathscr]{euscript}
\usepackage{bm}
\usepackage{textcomp}
\usepackage{graphicx}
\usepackage{subfigure}
\usepackage{overpic}
\usepackage{multirow}
\usepackage{floatrow}
\usepackage{float} 
\usepackage{amsmath,amssymb,amsthm}
\usepackage{cases}
\usepackage[colorlinks=true,linkcolor=red]{hyperref}
\usepackage{xcolor}
\newfloatcommand{capbtabbox}{table}[][\FBwidth]
\newcommand{\bea}{\begin{eqnarray}}
\newcommand{\eea}{\end{eqnarray}}
\newcommand{\beq}{\begin{equation}}
\newcommand{\eeq}{\end{equation}}

\def\/{\over}

\begin{document}

\title{Primordial black holes and scalar induced gravitational waves from sound speed resonance  in non-minimal derivative coupling inflation model}

\author{
Li-Shuai Wang$^{1}$\footnote{wlshuai212012@hainnu.edu.cn}, 
Qiong-Tao Xie$^{1}$\footnote{xieqiongtao@hainnu.edu.cn},
and Li-Yang Chen$^{2,3}$\footnote{lychen@cdnu.edu.cn (corresponding author)}
}

\affiliation{$^{1}$College of Physics and Electronic Engineering, Hainan Normal University, Haikou 571158, People's Republic of China\\$^{2}$College of Physics and Engineering Technology, \mbox{Chengdu Normal University, Chengdu, Sichuan 611130, People's Republic of China},\\ $^{3}$Department of Physics and Synergetic Innovation Center for Quantum Effects and Applications, \mbox{Hunan Normal University, Changsha, Hunan 410081, People's Republic of China} }

\begin{abstract}

We investigate an inflationary model with a non-minimal derivative coupling, where the coupling function contains both constant and periodic components. On large scales, the model is in excellent agreement with the latest Planck-ACT-LiteBIRD-BICEP/Keck 2018 (P-ACT-LB-BK18) observations. On small scales, the periodic component induces a sound-speed resonance mechanism that significantly amplifies curvature perturbations, resulting in the production of primordial black holes (PBHs).
By incorporating nonlinear effects in the PBH abundance calculation, we find that the resulting PBHs can account for the majority of dark matter in the Universe. Furthermore, the PBH formation process generates scalar-induced gravitational waves (SIGWs) with a characteristic multi-peak spectral shape, which may be detectable by future space-based detectors such as LISA, Taiji, and TianQin. The model also predicts a high-frequency stochastic gravitational-wave background (SGWB) from PBH binary mergers. 
A combined detection of SIGWs and high-frequency gravitational waves (GWs) in future experiments would provide a direct and testable probe of this inflationary scenario.

  \end{abstract}


\maketitle
\section{Introduction}
\label{sec_in}

During inflation, curvature perturbations are stretched beyond the Hubble radius and subsequently re-enter the horizon during the radiation- or matter-dominated eras. In regions with sufficiently large density contrasts, gravitational collapse can occur, leading to the formation of primordial black holes (PBHs)~\cite{Zeldovich, Hawking, Carr, Meszaros, Carr1975, Khlopov, Ozsoy2023}. PBHs have garnered significant attention due to their potential to explain various astrophysical and cosmological phenomena, including gravitational wave (GW) events observed by LIGO-Virgo~\cite{lg1, lg2, lg3, lg4}, ultrashort microlensing events in OGLE data~\cite{P.Mroz2017, H.Niikura2019}, and their possible role as dark matter candidates, particularly in the asteroid-mass range~\cite{A.Katz2018, A.Barnacka2012, P.W.Graham2015, H.Niikura2019a}. Moreover, the formation of PBHs is inherently tied to enhanced curvature perturbations, which can generate scalar induced gravitational waves (SIGWs) that may be detectable by future observatories~\cite{yfcai2018, yfcai2019}.
In order to generate a sizable population of PBHs, the amplitude of the curvature perturbation power spectrum must reach at least the order of $\mathcal{O}(10^{-2})$. Although the standard slow-roll inflation model is consistent with observations of the cosmic microwave background (CMB)~\cite{Aghanim2020}, it predicts an almost scale-invariant power spectrum with an amplitude of only $\mathcal{O}(10^{-9})$, which is insufficient to produce a significant population of PBHs. It is worth noting that while CMB data impose stringent constraints on large-scale perturbations, observational data on small-scale modes remain scarce, leaving room for the enhancement of perturbations at PBH-relevant scales.

To amplify primordial curvature perturbations during inflation, several mechanisms have been proposed. A commonly studied approach involves significantly slowing the inflaton’s motion ~\cite{Yokoyama1998, Choudhury2014, Germani2017, Motohashi2017, H. Di2018, Ballesteros2018, Dalianis2019, Gao2018, Tada2019, Mishra2020, Atal2020, Ragavendra2021, Bhaumik2020, Drees2021, C.Fu2020, Xu2020, Lin2020, Dalianis2021, Yi2021, Gao2021, Yi2021b, TGao2021, Solbi2021, Gao2021b, Solbi2021b, Zheng2021, Teimoori2021a, Cai2021, Wang2021, Fuchengjie2019, fuchengjie2020, Dalianis2020, Teimoori2021, Karam2022, Heydari2022, Heydari2022b, Bellido2017, Ezquiaga2018, Pi2022, Choudhury2023cc, Meng2022, Mu2022, Kawaguchi2022, Fu2022, Chen2022, Gu2023, Yi2022, S.Choudhury2023, Dalianis2023, Zhao2023, Heydari2023, Choudhury2023a, Firouzjahi2023, Mishra2023, Asadi2023, Cole2023, Choudhury2023b, Frolovsky2023, Chen2023a, Su2023, Cable2023, Escriva2023}. 
Alternatively, models with non-minimal derivative couplings increase gravitational friction, allowing slow-roll inflation to persist even with steep potentials, thereby enhancing perturbations~\cite{Fuchengjie2019, fuchengjie2020, Dalianis2020, Teimoori2021, Heydari2022, Heydari2022b}. 
Another class of mechanisms involves modifying the sound speed during inflation~\cite{O. Özsoy S. Parameswaran, G. Ballesteros, Kamenshchik,Gorji2022,Romano3,Ballesteros2022,R. Zhai,Qiu2022,Zhai2023}. In particular, sound speed resonance, where periodic variations in the sound speed induce parametric amplification of perturbations, has proven effective in generating enhanced small-scale curvature perturbations~\cite{G. Ballesteros, Kamenshchik, Gorji2022, Romano3, Ballesteros2022, R. Zhai, Qiu2022, Zhai2023, yfcai2018, yfcai2019, c.chen2019, c.chen2020, Addazi2022, bLi2023, lyc2024}. Another similar mechanism is parametric resonance, where periodic features in the potential drive the enhancement~\cite{ Cai2020}.
Recently, the incorporation of a ultra-slow-roll phase within curvaton inflation models has been shown to significantly enhance small-scale curvature perturbations, leading to PBH formation~\cite{2502.14323}. Hybrid inflation scenarios driven by axion-like particles have also been studied, suggesting a PBH mass range tightly correlated with the second-order GW spectrum~\cite{2411.00764}. The Type III hilltop inflation model predicts amplified small-scale perturbations, though their amplitude may still be insufficient for efficient PBH production~\cite{2407.04443}. These findings, along with further studies on transient slow-roll violations~\cite{2405.12149}, multi-phase inflation~\cite{2404.02492}, spectator fields in supergravity~\cite{2404.02295}, and modifications to dark energy or kinetic structures~\cite{2401.15451, 2401.10925, 2307.08046}, provide deeper insights into inflationary dynamics favorable for PBH formation.

Motivated by the above considerations, we investigate an inflationary model with a non-minimal derivative coupling, in which the coupling function consists of both a constant term and a periodic term. This structure, while maintaining consistency with the latest P-ACT-LB-BK18 constraints~\cite{A. Adame et al., A. Adame et al2.} at large scales, can excite oscillations in the sound speed at small scales, thereby triggering a sound-speed resonance mechanism that amplifies curvature perturbations in specific modes. We systematically analyze the enhancement of perturbations induced by sound-speed resonance and incorporate nonlinear corrections to evaluate the abundance of PBHs. Furthermore, the SIGWs generated during the PBH formation process are also within the scope of our study. We analyze their spectral characteristics and assess their potential detectability by future space-based GW observatories such as Taiji~\cite{taiji}, TianQin~\cite{tianqin}, and LISA~\cite{lisa}. In addition, we investigate the high-frequency stochastic gravitational-wave background (SGWB) produced by PBH binary mergers. Our preliminary analysis suggests that, due to the extremely high frequencies associated with low-mass black holes, such signals are unlikely to be directly detectable by current detectors. However, if they can be jointly observed with SIGWs in the future, we anticipate that such multi-signal observations would provide a powerful test of this inflationary scenario.

The structure of this paper is as follows: Section~\ref{sec2} presents the inflationary model with non-minimal derivative coupling. Section~\ref{sec3} investigates sound speed resonance and its role in enhancing curvature perturbations. Section~\ref{sec4} examines PBH formation, SIGW generation, and the SGWB from binary PBH mergers. Section~\ref{conclusion} concludes the paper.

\section{Inflation model with derivative coupling}
\label{sec2}

We consider a single scalar field inflation model in which the inflaton field $\phi$ couples derivatively with the Einstein tensor $G_{\mu\nu}$. The system is described by the following action
\begin{align}\label{action1}
\mathcal{S}=\int d^{4}x \sqrt{-g}\left[\frac{M_{\mathrm{pl}}^2}{2} R-\frac{1}{2}\left(g^{\mu \nu}-\frac{1}{M_{\mathrm{pl}}^2} \theta(\phi) G^{\mu \nu}\right) \nabla_{\mu} \phi \nabla_{\nu} \phi - V(\phi)\right],
\end{align}
where  $g$ is the determinant of the metric tensor $g_{\mu\nu}$, $M_{\mathrm{pl}}$ is the reduced Planck mass, $R$ is the Ricci scalar, $\theta(\phi)$ is the dimensionless coupling function, and $V(\phi)$ is the potential of the inflaton field.

In the spatially flat Friedmann-Robertson-Walker background, the equations of motion derived from the action (\ref{action1}) are given by
\begin{align}\label{br1}
3 H^{2}=\frac{1}{M_{\mathrm{pl}}^{2}}\left[\frac{1}{2}\left(1+ \frac{9}{M_{\mathrm{pl}}^{2}} \theta(\phi) H^{2} \right) \dot{\phi}^{2}+V(\phi)\right],
\end{align}

\begin{align}\label{br2}
-2 \dot{H}=\frac{1}{M_{\mathrm{pl}}^{2}}\left[\left(1+\frac{3}{M_{\mathrm{pl}}^{2}} \theta(\phi) H^{2}-\frac{1}{M_{\mathrm{pl}}^{2}} \theta(\phi) \dot{H}\right) \dot{\phi}^{2}- \frac{1}{M_{\mathrm{pl}}^{2}} \theta_{,\phi}(\phi) H \dot{\phi}^{3} - \frac{2}{M_{\mathrm{pl}}^{2}} \theta(\phi) H \dot{\phi} \ddot{\phi}\right],
\end{align}

\begin{align}\label{br3}
\left(1+ \frac{3}{M_{\mathrm{pl}}^{2}} \theta(\phi) H^{2}\right) \ddot{\phi}+\left[1+\frac{1}{M_{\mathrm{pl}}^{2}} \theta(\phi)\left(2 \dot{H}+3 H^{2}\right)\right] 3 H \dot{\phi} + \frac{3}{2M_{\mathrm{pl}}^{2}} \theta_{,\phi} H^{2} \dot{\phi}^{2}+V_{,\phi}=0.
\end{align}
Here, $H$ is the Hubble parameter, an overdot denotes a derivative with respect to cosmic time $t$, $\theta_{,\phi} \equiv d\theta/d\phi$, and $V_{,\phi} \equiv dV/d\phi$. To describe the slow-roll inflation, we define four slow-roll parameters
\begin{align}\label{SRP}
\epsilon =-\frac{\dot{H}}{H^{2}}, \quad \delta_{\phi} =\frac{\ddot{\phi}}{H \dot{\phi}}, \quad 
\delta_{X} =\frac{\dot{\phi}^{2}}{2 M_{\mathrm{pl}}^{2} H^{2}}, \quad \delta_{D} =\frac{\theta \left(\phi \right) \dot{\phi}^{2}}{4 M_{\mathrm{pl}}^{4}}.
\end{align}
The slow-roll condition is satisfied if ${\epsilon, |\delta_{\phi}|, |\delta_{X}|, \delta_{D}} \ll 1$.

During inflation, the quantum fluctuations of the inflaton field and the metric serve as the seeds for the large-scale structure of the universe. Scalar fluctuations are typically described by the curvature perturbations $\mathcal{R}$. By expanding the action (\ref{action1}) to second order in $\mathcal{R}$~\cite{A.D.Felice2011, S.Tsujikawa2012, Kobayashi2011}, we obtain
\begin{align}
S^{(2)}=\int d t d^{3} x a^{3} Q\left[\dot{\mathcal{R}}^{2}-\frac{c_{s}^{2}}{a^{2}}(\partial \mathcal{R})^{2}\right],
\end{align}
where 
\begin{align}\label{Q}
Q=\frac{w_{1}\left(4 w_{1} w_{3}+9 w_{2}^{2}\right)}{3 w_{2}^{2}},
\end{align}
and
\begin{align}\label{cs}
c_{s}^{2}=\frac{3\left(2 w_{1}^{2} w_{2} H-w_{2}^{2} w_{4}+4 w_{1} \dot{w}_{1} w_{2}-2 w_{1}^{2} \dot{w}_{2}\right)}{w_{1}\left(4 w_{1} w_{3}+9 w_{2}^{2}\right)}.
\end{align}
Here, $c_{s}^{2}$ is the square of the sound speed of scalar perturbations. The coefficients $w_i$ are defined as
\begin{align}\label{wi}
w_{1}&=M_{\mathrm{pl}}^{2}\left(1-2 \delta_{D}\right), \nonumber \\
w_{2}&=2 H M_{\mathrm{pl}}^{2}\left(1-6 \delta_{D}\right), \nonumber \\
w_{3}&=-3 H^{2} M_{\mathrm{pl}}^{2}\left(3-\delta_{X}-36 \delta_{D}\right), \nonumber \\
w_{4}&=M_{\mathrm{pl}}^{2}\left(1+2 \delta_{D}\right).
\end{align}
Defining $u_{k} = z\mathcal{R}_{k}$ with $z \equiv a\sqrt{2Q}$, the evolution of $u_{k}$ is 
\begin{align}\label{sasaki2}
\ddot{u}_k + H\dot{u}_k + \left( \frac{c_{s}^{2}k^{2}}{a^{2}} -\frac{\ddot{z}+H\dot{z}}{z} \right) u_k = 0,
\end{align}
and the power spectrum of curvature perturbations is given by
\begin{align}\label{prk}
\mathcal{P}_\mathcal{R}(k) = \frac{k^{3}}{2 \pi^{2}} \left|\frac{u_{k}}{z}\right|^{2}.
\end{align}
Using the definitions of $z$ and $u_{k}$, the power spectrum can be rewritten as~\cite{S.Tsujikawa2012}
\begin{align}
\mathcal{P}_{\mathcal{R}} \simeq \frac{V^3}{12 \pi^2 M_{\mathrm{pl}}^6 V_{,\phi}^2}\left(1+\theta(\phi) \frac{V}{M_{\mathrm{pl}}^4}\right) \equiv \mathcal{P}_{\mathcal{R}_0} \mathcal{A},
\end{align}
where  
\begin{align}
\mathcal{A}=1+\theta(\phi) \frac{V}{M_{\mathrm{pl}}^4},
\end{align}
and $\mathcal{P}_{\mathcal{R}_0}$ is the power spectrum in the minimal coupling case. Thus the spectral index $n_s$ and tensor-to-scalar ratio 
$r$ are then expressed as
\begin{align}
n_s \simeq 1-\frac{1}{\mathcal{A}}\left[2 \epsilon_V\left(4-\frac{1}{\mathcal{A}}\right)-2 \eta_V\right],
\end{align}
\begin{align}
r \simeq \frac{16 \epsilon_V}{\mathcal{A}},
\end{align}
where $\epsilon_V = \frac{1}{2}M_{\mathrm{pl}}^2 \left(\frac{V_{,\phi}}{V}\right)^2$ and $\eta_V = M_{\mathrm{pl}}^2 \frac{V_{,\phi\phi}}{V}$ are slow-roll parameters defined from the inflaton potential.

\section{Resonant Amplification of Curvature Perturbations}
\label{sec3}

When superhorizon curvature perturbations re-enter the Hubble horizon during the radiation- or matter-dominated eras, they can collapse overdense regions, potentially leading to the formation of PBHs. However, within the framework of standard slow-roll inflation, the probability of such collapses is exceedingly small due to the insufficient amplitude of the curvature perturbation power spectrum. To generate a substantial abundance of PBHs, the amplitude of the power spectrum must be enhanced by a minimum of approximately seven orders of magnitude. In this section, we investigate how a small, periodic derivative coupling can induce a resonant amplification of curvature perturbations, thereby achieving the required enhancement.

In this study, we adopt an inflationary potential of the power-law form~\cite{Linde}
\begin{align}\label{Potential}
V(\phi) = \Lambda\, \phi^{2/5},
\end{align}
where $\Lambda$ is a constant parameter.  
The coupling function $\theta(\phi)$ is defined as
\begin{align}\label{theta}
\theta(\phi) = m + w \sin\!\left(n\phi\right) 
\Theta\!\left(\phi_{s} - \phi\right) \Theta\!\left(\phi - \phi_{e}\right),
\end{align}
Here, \( m \) is a constant, and \( w \) is a small positive constant satisfying \( w \ll m \), ensuring that the oscillatory component of the coupling has a negligible impact on the overall background dynamics. The parameter \( n \), with mass dimension \( 1/M_{\mathrm{pl}} \), controls the oscillation frequency. When \( n \gg \phi^{-1} \), the coupling function exhibits pronounced oscillatory behavior within the interval \( \phi_e < \phi < \phi_s \). Here, \( \Theta \) denotes the Heaviside step function, which restricts the oscillations to a finite window during inflation.  
In this setup, \( \phi_s \) is chosen to be significantly smaller than the field value at the onset of inflation, thereby ensuring that the enhanced perturbations are generated exclusively on small scales.

Under the slow-roll conditions given in Eq.~(\ref{SRP}), along with \( w \ll m \) and \( n \gg \phi^{-1} \), the sound speed of curvature perturbations in Eq.~(\ref{cs}) can be approximated as
\begin{align}\label{cs2}
c_s^2 \approx 1 + \delta c_s = 1 + \frac{3 \dot{\phi}_s^3 H_s m n w \cos (n \phi)}{M_{\mathrm{pl}}^6 \left(1 + \frac{3}{M_{\mathrm{pl}}^2} H_s^2 m\right)^2}.
\end{align}
Here, the term \( \delta c_s \), which satisfies \( |\delta c_s| \ll 1 \), oscillates around zero. 
The oscillatory sound speed can resonantly amplify curvature perturbations. This resonance occurs deep inside the horizon, where \( c_s k \gg aH \). Under this condition, Eq.~(\ref{sasaki2}) simplifies to
\begin{align}\label{sasaki3}
\ddot{u}_{k} + c_{s}^{2} \frac{k^{2}}{a^{2}} u_{k} \approx 0.
\end{align}
Substituting Eq.~(\ref{cs2}) into Eq.~(\ref{sasaki3}), we obtain
\begin{align}\label{madiu3}
\ddot{u}_k + \left(\frac{k^2}{a^2} + \frac{3 \dot{\phi}_s^3 H_s m n w \cos (n \phi) k^2}{M_{\mathrm{pl}}^6 \left(1 + \frac{3}{M_{\mathrm{pl}}^2} H_s^2 m\right)^2 a^2}\right) u_k \approx 0.
\end{align}
Since the time interval for the inflaton field to evolve from \( \phi_s \) to \( \phi_e \) is very short, we approximate \( \phi \approx \phi_s + \dot{\phi}_s (t - t_s) \), where \( t_s \) is the time when \( \phi = \phi_s \) ($t_{e}$ is the moment when  $\phi = \phi_e$ in the following discussion ). Defining \( 2x = n \dot{\phi}_s t + n (\phi_s - \dot{\phi}_s t_s) \) and \( k_n^2 = \frac{n^2 \dot{\phi}_s^2}{4} \), Eq.~(\ref{madiu3}) can be rewritten as a Mathieu equation
\begin{align}\label{madiu0}
\frac{d^{2} u_{k}}{d x^{2}} + \left[A_{k}(x) - 2q \cos 2x\right] u_{k} = 0,
\end{align}
where
\begin{align}\label{madiu1}
A_k(x) \equiv \frac{k^2}{k_n^2 a^2}, \quad q \equiv -\frac{6 H_s \dot{\phi}_s m w k^2}{M_{\mathrm{pl}}^6 n a^2 \left(1 + \frac{3}{M_{\mathrm{pl}}^2} H_s^2 m\right)^2} = 2C \frac{k^2}{k_n^2 a^2}.
\end{align}

The solutions to the Mathieu equation are known as Mathieu functions, which exhibit two distinct types of resonance: broad resonance (\( q > 1 \)) and narrow resonance (\( 0 < q \ll 1 \)). In this paper, we focus on the case of narrow resonance. Resonance occurs when \( A_k \approx \alpha^2 \) (\( \alpha = 1, 2, 3, \dots \)). Among these, the oscillation is most pronounced for \( \alpha = 1 \), so we concentrate on the first resonance band. The width of each resonance band is approximately \( \Delta k \sim q^{\alpha} \). In the regime of narrow resonance, the Mathieu function grows exponentially as \( u_k \propto \exp(\mu_k x) \), where the growth rate \( \mu_k(t) \) is given by
\begin{align}\label{mu}
\mu_k(t) = \sqrt{\left(\frac{q}{2}\right)^2 - \left(\frac{k}{k_n a} - 1\right)^2} = \sqrt{\left(C \frac{k^2}{k_n^2 a^2}\right)^2 - \left(\frac{k}{k_n a} - 1\right)^2}.
\end{align}
For resonance to occur, the term inside the square root must remain positive, leading to the condition
\begin{align}\label{k}
k_- < \frac{k}{a} < k_+,
\end{align}
where \( k_- = k_n (1 - |C|) \) and \( k_+ = k_n (1 + |C|) \). Since \( q \) satisfies \( 0 < q \ll 1 \), it follows that \( |C| \ll 1 \). The time dependence of Eq.~(\ref{k}) ensures that the \( k \)-mode remains in the resonant band only for a finite duration, given by 
\[
T_{\text{in}}(k) = \min(t_e, t_F) - \max(t_s, t_I),
\]
where \( t_I \) and \( t_F \) are the times when the \( k \)-mode enters and exits the resonant band, respectively.
The resonant amplification width of the power spectrum, \( \Delta k = k_+ a_e - k_- a_s \), primarily depends on the parameters \( k_n \), \( \phi_s \) and \( \phi_e \). Here $a_{s}$ and $a_{e}$ denote the values of scale factor $a$ at cosmic times $t=t_{s}$ and $t= t_{e}$, respectively.

During sound speed resonance, curvature perturbations experience exponential amplification, given by
\begin{align}\label{E}
\mathcal{E}(k) = \frac{u_k(t_F)}{u_k(t_I)} \approx \exp \left( \int_{t_I}^{t_F} \mu_k(t) k_n \, dt \right).
\end{align}  
Defining \( \mathcal{B}_k(t) \equiv \frac{k}{k_n a} \), Eq.~(\ref{E}) can be rewritten as  
\begin{align}
\mathcal{E}_k\left(\mathcal{B}_k(t_I), \mathcal{B}_k(t_F)\right) \approx \exp \left(-\frac{k_n}{H_s} \int_{\mathcal{B}_k(t_I)}^{\mathcal{B}_k(t_F)} \sqrt{\left(C \mathcal{B}_k\right)^2 - \left(\mathcal{B}_k - 1\right)^2} \frac{d \mathcal{B}_k}{\mathcal{B}_k} \right).
\end{align}  
The amplified modes can be classified into three groups: (1) the modes enter the resonant band before \( t_s \); (2)the modes enter the resonant band after \( t_s \) and exit before \( t_e \); (3)the modes exit the resonant band after \( t_e \). For these groups, the wavenumbers satisfy \( k_{-} a_s < k \leq k_{+} a_s \), \( k_{+} a_s < k < k_{-} a_e \), and \( k_{-} a_e \leq k < k_{+} a_e \), respectively. We can computed \( \mathcal{B}_k(t_I) \) and \( \mathcal{B}_k(t_F) \) as follow
\begin{eqnarray}
	\mathcal{B}_{k}(t_I)=\frac{k}{k_n a_s}, \quad \mathcal{B}_{k}(t_F)=\frac{k_-}{k_n} \quad \mathrm{for}\quad k_- a_s <k\leq k_+ a_s, 
\end{eqnarray}
\begin{equation}\label{not k}
	\mathcal{B}_k\left(t_I\right)=\frac{k_{+}}{k_n}, \quad \mathcal{B}_k\left(t_F\right)=\frac{k_{-}}{k_n} \quad \text { for } \quad k_{+} a_s<k<k_{-} a_e,
\end{equation}
\begin{equation}
	\mathcal{B}_k\left(t_I\right)=\frac{k_{+}}{k_n}, \quad \mathcal{B}_k\left(t_F\right)=\frac{k}{k_n a_e} \quad \text { for } \quad k_{-} a_e \leq k<k_{+} a_e.
\end{equation}
In second group, $\mathcal{B}_{k}(t_I)$ and $\mathcal{B}_{k}(t_F)$ are independent of $k$, so \( \mathcal{E}_k \) is independent of $k$ too. For all cases, the enhanced power spectrum of curvature perturbations can be expressed as
\begin{align}\label{red mathieu}
\mathcal{P}_{\mathcal{R}}(k) \approx \mathcal{E}_k^2 \mathcal{P}_{\mathcal{R}_0}(k).
\end{align}

\begin{figure*}[h!]
    \centering
    \includegraphics[width=0.55\linewidth]{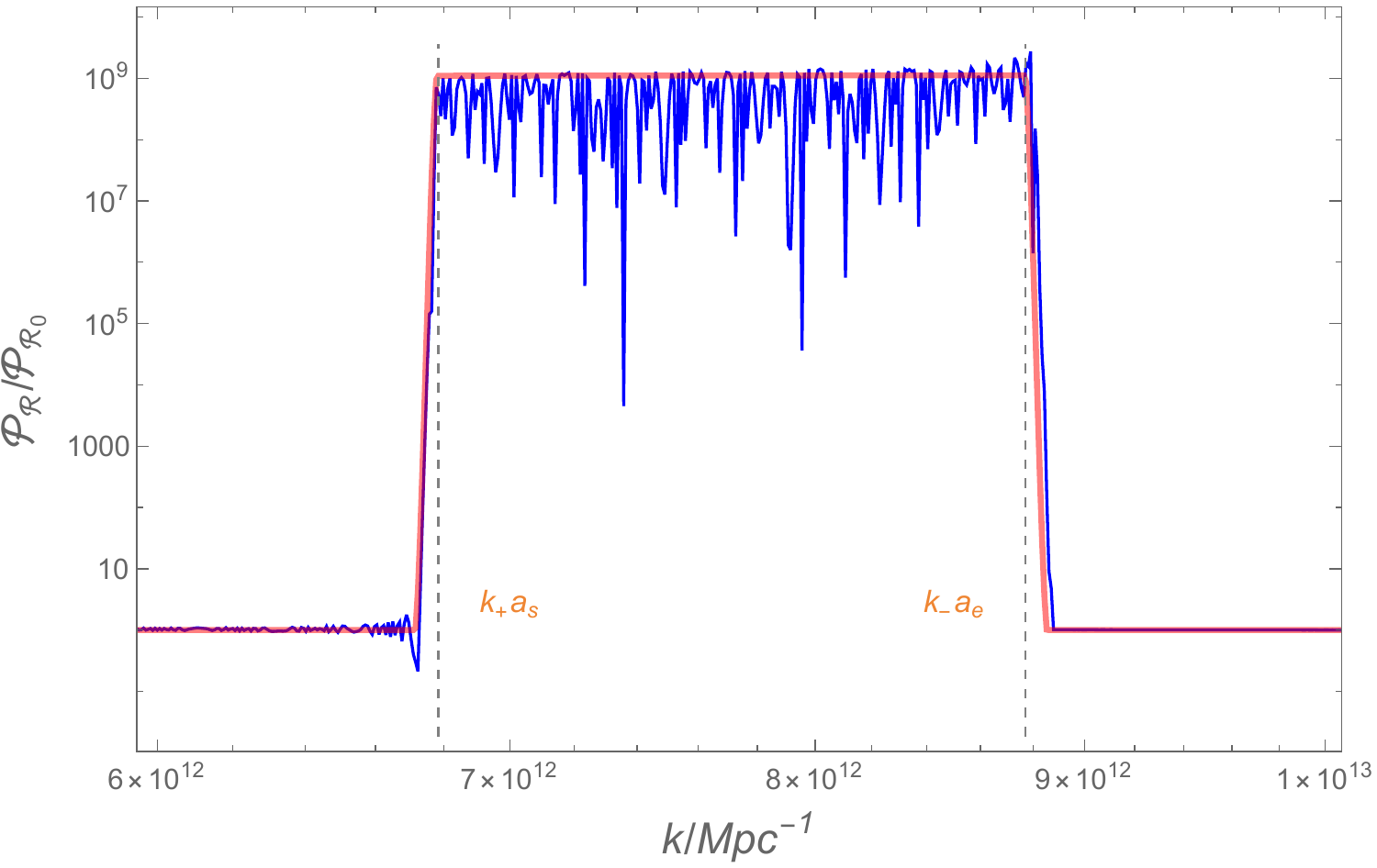}
    \caption{\label{AK}
        The normalized power spectrum enhancement $\mathcal{P}_\mathcal{R}/\mathcal{P}_{\mathcal{R}_0}$ versus $k/Mpc^{-1}$. Numerical solutions (blue) from $\mathcal{P}_\mathcal{R}/\mathcal{P}_{\mathcal{R}_0}=\frac{k^{3}}{2 \pi^{2}} \left|\frac{u_{k}}{z}\right|^{2}/\mathcal{P}_{\mathcal{R}_0}$ demonstrate excellent agreement with analytical predictions (red) from Eq.~(\ref{red mathieu}), showing the characteristic plateau feature predicted by the resonant amplification mechanism.
    }
\end{figure*}

\begin{figure*}
    \centering
    \includegraphics[width=0.55\linewidth]{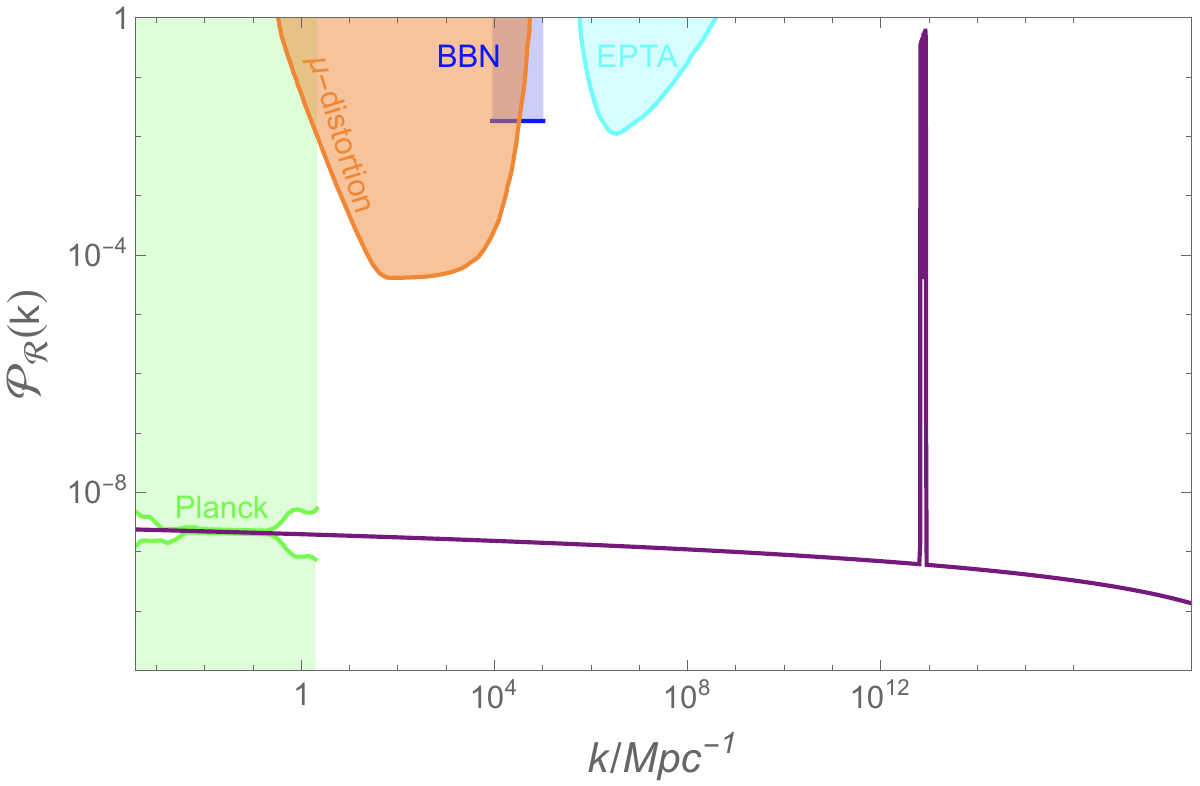}
    \caption{\label{PR}
        The primordial curvature perturbation power spectrum (solid purple line) with observational bounds. The spectrum satisfies constraints from CMB (green)~\cite{Aghanim2020}, $\mu$ distortion (orange)~\cite{D.J.Fixsen}, BBN (blue)~\cite{K.Inomata2016}, and EPTA observations (cyan)~\cite{K.Inomata2019}, while achieving sufficient enhancement at small scales for PBH formation.
    }
\end{figure*}

With parameters set to \( m = 9.0\times10^{8} \), \( w = 1.26472\times10^{5} \), \( \Lambda = 4.93 \times10^{-10} M_\mathrm{pl}^{10} \), \( \phi_{s} = 3.798 M_\mathrm{pl} \), \( \phi_{e} = 3.782 M_\mathrm{pl} \), \( n = 1\times10^{7} M_\mathrm{pl}^{-1} \), and e-folding number \( N \simeq 50 \), our model predicts the tensor-to-scalar ratio \( r \simeq 0.03 \) and the spectral index $n_s\simeq0.9758$, which are in excellent agreement with current observations from P-ACT-LB-BK18.
Figure~\ref{AK} illustrates the amplification of the curvature perturbation power spectrum in the resonant region \( k_{+} a_s < k < k_{-} a_e \). The blue curve represents the numerical solution derived from $\mathcal{P}_\mathcal{R}/\mathcal{P}_{\mathcal{R}_0}=\frac{k^{3}}{2 \pi^{2}}\left|\frac{u_{k}}{z}\right|^{2}/\mathcal{P}_{\mathcal{R}_0}$, while the red curve corresponds to the analytical approximation given by Eq.~(\ref{red mathieu}). As anticipated from the analysis in Eq.~(\ref{not k}), this amplification produces a characteristic plateau in this region, indicating that the amplification factor is independent of the wavenumber \( k \). The excellent agreement between numerical and analytical results validates the reliability of our theoretical framework and demonstrates that this mechanism can enhance the power spectrum by several orders of magnitude.
Figure~\ref{PR} presents the complete power spectrum of curvature perturbations obtained by numerically solving Eq.~(\ref{prk}). Our results are consistent with CMB observations on large scales, while successfully achieving significant enhancement of perturbations on small scales.

\section{PBHs and GWs}\label{sec4}

During inflation, curvature perturbations are stretched beyond the Hubble horizon and subsequently re-enter during the radiation- or matter-dominated eras. Perturbations with sufficiently large amplitudes can induce significant density contrasts upon horizon re-entry, potentially triggering gravitational collapse in over-dense regions and leading to the formation of PBHs.

The mass of a PBH is related to the horizon mass at the time of collapse~\cite{Lorenzo Frosina} by the expression
\begin{align}
M_{\rm PBH} = \mathcal{K} M_H \left[\left(\delta_{\rm L} - \frac{1}{4\Phi}\delta_{\rm L}^2\right) - \delta_{\textrm{th}}\right]^{\gamma}\,,
\end{align}
where the constants are set to \( \mathcal{K} = 4.36 \), \( \gamma = 0.38 \), and \( \Phi = 2/3 \). Here, \( \delta_{\rm L} \) denotes the linear Gaussian component of the density contrast, and the critical threshold for gravitational collapse is given by \( \delta_{\textrm{th}} \simeq 0.594 \), which is obtained from our calculation based on the curvature power spectrum considered in this work~\cite{Lorenzo Frosina,D. Frolovsky}. The horizon mass \( M_H \) is related to the comoving wavenumber \( k \) by~\cite{Lorenzo Frosina}
\begin{align}
M_H\simeq 17 \left(\frac{g_*}{10.75}\right)^{-1/6} \left(\frac{k}{10^6\, \mathrm{Mpc}^{-1}}\right)^{-2}\, M_{\odot}\,,
\end{align}
where \( g_* = 106.75 \) is the effective number of relativistic degrees of freedom during the early radiation-dominated era, and \( M_{\odot} \) is the solar mass.

The fraction of the Universe's energy density collapsing into PBHs of horizon mass \( M_H \) is given by
\begin{align}
\beta(M_H) & = 
\int_{\delta_{\textrm{th}}}^{\infty}\frac{M_{\rm PBH}}{M_H}P(\delta)\,d\delta \,, \label{eq:BetaDef}\\
& = 
\mathcal{K}
\int_{\delta_{\rm L}^{\rm min}}^{\delta_{\rm L}^{\rm max}}
\left[\left(\delta_{\rm L} - \frac{1}{4\Phi}\delta_{\rm L}^2\right) - \delta_{\textrm{th}}\right]^{\gamma}
P_{\rm G}(\delta_{\rm L})\,d\delta_{\rm L}\,,
\end{align}
where the integration limits are defined as
\begin{align}
\delta_{\rm L}^{\rm min} = 2\Phi\left(
1 - \sqrt{1-\frac{\delta_{\textrm{th}}}{\Phi}}
\right)\,, \qquad \delta_{\rm L}^{\rm max} = 2\Phi\,,
\end{align}
and \( P_{\rm G}(\delta_{\rm L}) \) denotes the Gaussian probability distribution of the linear density contrast
\begin{align}
P_{\rm G}(\delta_{\rm L}) = \frac{1}{\sqrt{2\pi}\sigma(M_H)}\,\exp\left(-\frac{\delta_{\rm L}^2}{2\sigma^2(M_H)}\right)\,.
\end{align}
The variance \( \sigma^2(M_H) \) of the coarse-grained density contrast, smoothed over the scale \( k \), is given by~\cite{S. Wang}
\begin{equation}
\sigma^2(M_H) = \frac{4}{9} \Phi^{2}\int d \ln q \, W^2\left(q k^{-1}\right) \left(q k^{-1}\right)^4 \mathcal{P}_{\mathcal{R}}(q)\,\mathcal{T}^{2} (qk^{-1})\,,
\end{equation}
where \( W(y) = 3\left[\frac{\sin y - y\cos y}{y^3}\right] \) is the Fourier transform of the real-space top-hat window function, and
\begin{align}
\mathcal{T}(y) \equiv 3\left[
\frac{\sin(c_s y) - (c_s y)\cos(c_s y)}{(c_s y)^3}
\right]
\end{align}
is the linear transfer function in the radiation-dominated epoch. The sound speed is \( c_s = 1/\sqrt{3} \) in this era.

The current PBH abundance as a fraction of the total dark matter energy density is expressed as
\begin{equation}\label{PBHs/DM}
f_{\rm PBH}(M_{\rm PBH})= \int \frac{d M_{\rm PBH}}{M_{\rm PBH}} f(M_{\rm PBH}),
\end{equation}
where the differential mass function \( f(M_{\rm PBH}) \) is given by~\cite{C. T. Byrnes, G. Franciolini and  A. Urbano}
\begin{align}\label{fm}
&f(M_{\rm PBH}) = \frac{1}{\Omega_{\rm CDM}}\int_{M_H^{\rm min}}^{\infty}
\left(
\frac{M_{\rm eq}}{M_H}
\right)^{1/2} 
\frac{
e^{-\frac{8}{9\sigma^2(M_H)}\left[
1 - 
\sqrt{\Lambda}\,
\right]
^2}
}{
\sqrt{2\pi}\sigma(M_H)
\Lambda^{1/2}}\left(
\frac{M_{\rm PBH}}{\gamma M_H}
\right)\left(
\frac{M_{\rm PBH}}{\mathcal{K}M_H}
\right)^{1/\gamma}d\ln M_H
\,,
\end{align}
with
\begin{equation}
\Lambda \equiv 1 - \frac{\delta_{\textrm{th}}}{\Phi} - \frac{1}{\Phi}\left(\frac{M_{\rm PBH}}{\mathcal{K}M_H}\right)^{1/\gamma}\,,
\end{equation}
and the \( M_{\rm eq} \simeq 3 \times 10^{17}\, M_{\odot} \) denoting the horizon mass at matter-radiation equality.

\begin{figure}[h]
\centering
\includegraphics[width=0.55\linewidth]{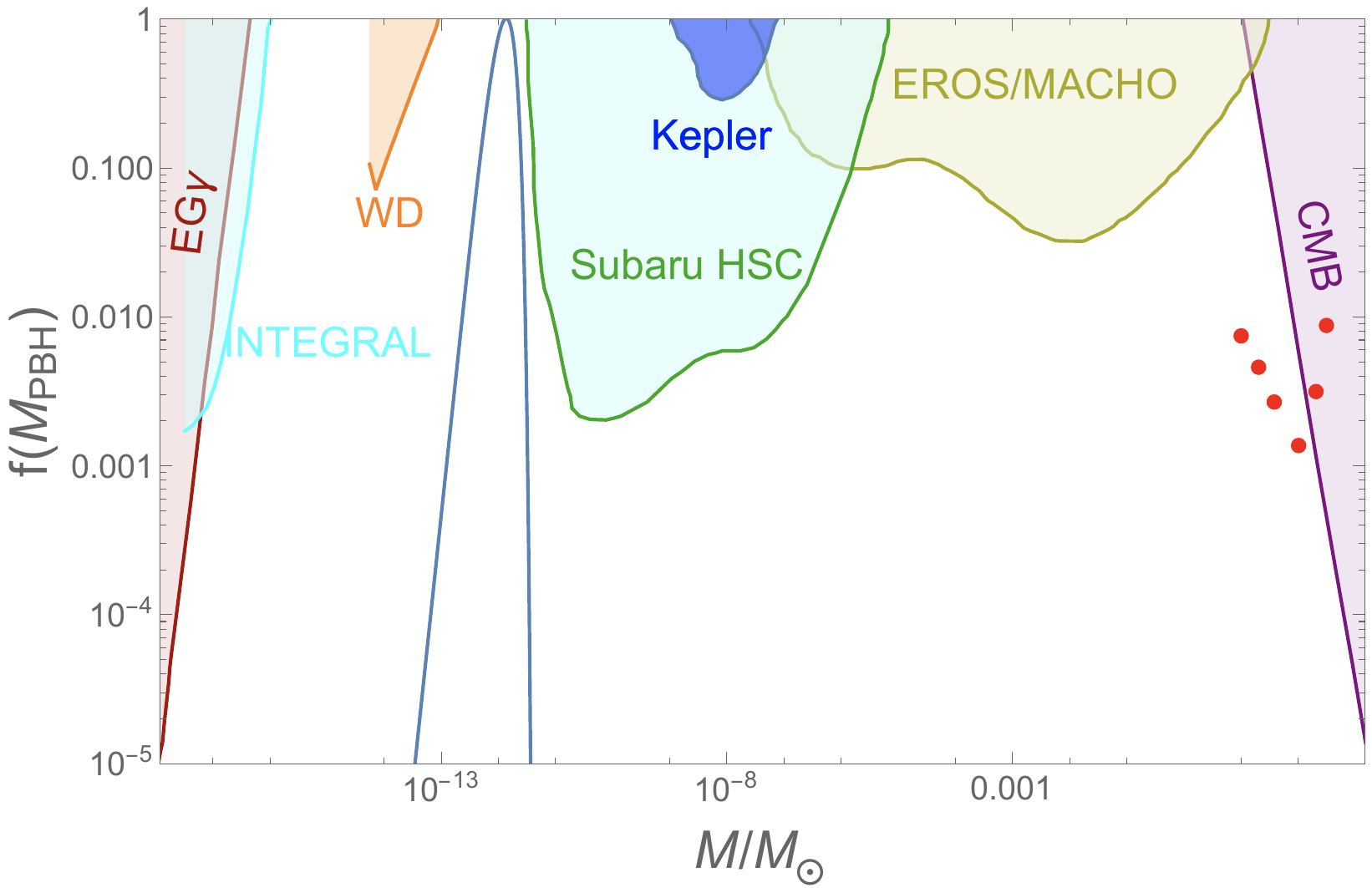}
\caption{\label{PBH} Predicted PBH mass spectrum (solid light-blue line) in comparison with current observational constraints, including bounds from EG$\gamma$~\cite{B.J.Carr2010}, white dwarfs (WD)~\cite{P.W.Graham2015}, INTEGRAL~\cite{R.Laha}, Kepler~\cite{K.Griest}, Subaru HSC~\cite{H.Niikura2019}, EROS/MACHO~\cite{P.Tisserand}, and CMB~\cite{V. Poulin}.}
\end{figure}

Based on the curvature power spectrum \( \mathcal{P}_\mathcal{R}(k) \) obtained in the previous section, our numerical evaluation of Eqs.~(\ref{PBHs/DM}) and (\ref{fm}) suggests that PBHs with a peak mass around \( 10^{-12} M_{\odot} \) can account for nearly the entirety of the cold dark matter, yielding \( f_{\rm PBH}   \approx 0.99 \). The resulting PBH mass spectrum is illustrated in Figure~\ref{PBH}, alongside existing observational constraints.

\begin{figure*}
\centering
\includegraphics[width=0.55\linewidth]{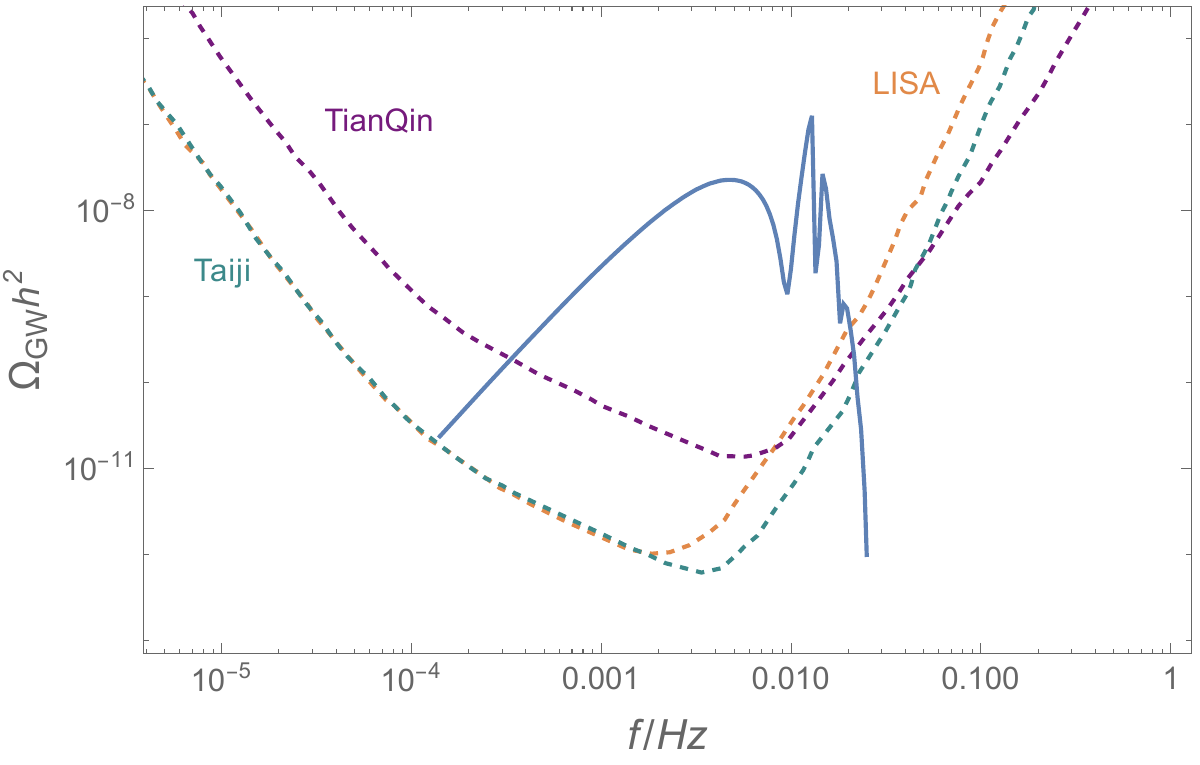}
\caption{\label{GW} Present-day SIGW energy spectrum (solid blue line) compared with the sensitivity curves (dotted lines) of GW detectors TAIJI~\cite{taiji}, TIANQIN~\cite{tianqin}, and LISA~\cite{lisa}.}
\end{figure*}

\begin{figure*}
\centering
\includegraphics[width=0.55\linewidth]{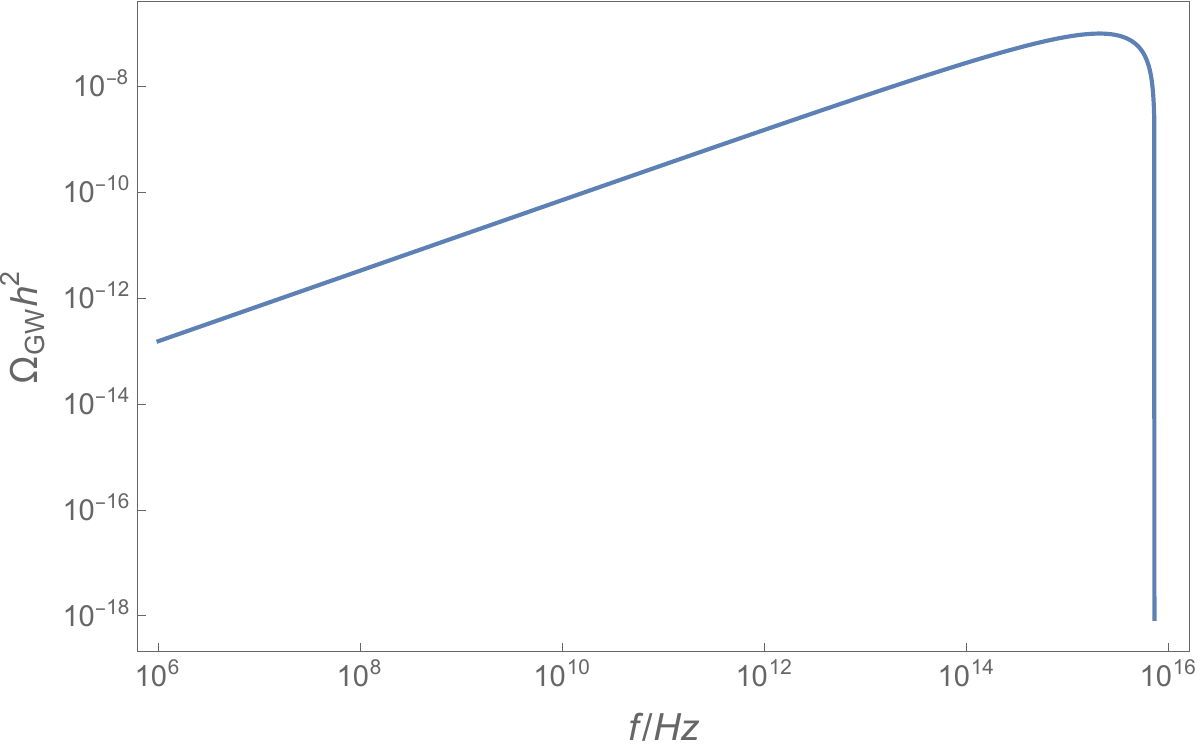}
\caption{\label{SGWB} The SGWB signal is produced by binary PBH mergers with masses around \( 10^{-12} M_{\odot} \), where these PBHs constitute nearly all dark matter.}
\end{figure*}

In parallel, SIGWs are generated during PBH formation through second-order tensor perturbations sourced by large curvature perturbations. The evolution of tensor perturbations $h_{ij}$ is governed by \cite{Baumann2007,Ananda2007,J. R. Espinosa,K. Kohri,arXiv.2309.03101,Yogesh Abolhassan Mohammadi2025}
\begin{equation}\label{GWS}
h_{i j}^{\prime \prime}+2 \mathcal{H} h_{i j}^{\prime}-\nabla^2 h_{i j}=-4 \mathcal{T}_{i j}^{l m} S_{l m},
\end{equation}
where $\mathcal{H} \equiv a'/a$ is the conformal Hubble parameter, a prime represents differentiation with respect to conformal time, $\mathcal{T}_{i j}^{l m}$ is a projection operator, and $S_{ij}$ is the source term involving scalar perturbations $\Psi$
\begin{align}
S_{ij} = 4 \Psi \partial_i \partial_j \Psi + 2 \partial_i \Psi \partial_j \Psi - \frac{1}{\mathcal{H}^2} \partial_i\left(\mathcal{H} \Psi+\Psi^{\prime}\right) \partial_j\left(\mathcal{H} \Psi+\Psi^{\prime}\right).
\end{align}
During the radiation-dominated era ($w = 1/3$), the scalar perturbations $\Psi$ evolve as
\begin{equation}
\Psi_k(\eta) = \psi_k \frac{9}{(k \eta)^2} \left(\frac{\sin(k \eta / \sqrt{3})}{k \eta / \sqrt{3}} - \cos(k \eta / \sqrt{3})\right),
\end{equation}
where $\psi_k$ is linked to the primordial curvature power spectrum $\mathcal{P}_{\mathcal{R}}(k)$. The GW energy density is computed as \cite{Kohri2018}
\begin{align}\label{gwk}
\Omega_{\mathrm{GW}}\left(\eta_c, k\right) &= \frac{1}{12} \int_{0}^{\infty} d v \int_{|1-v|}^{|1+v|} d u \left(\frac{4 v^2 - \left(1 + v^2 - u^2\right)^2}{4 u v}\right)^2 \mathcal{P}_{\mathcal{R}}(k u) \mathcal{P}_{\mathcal{R}}(k v) \nonumber\\
&\quad \times \left(\frac{3}{4 u^3 v^3}\right)^2 \left(u^2 + v^2 - 3\right)^2 \nonumber\\
&\quad \times \left\{\left[-4 u v + \left(u^2+v^2-3\right) \ln \left|\frac{3-(u+v)^2}{3-(u-v)^2}\right|\right]^2 \right. \nonumber\\
&\quad \left. + \pi^2 \left(u^2+v^2-3\right)^2 \Theta(v+u-\sqrt{3})\right\},
\end{align}
where $\eta_c$ denotes the conformal time when SIGW production ceases. The present-day SIGW energy density spectrum is
\begin{align}\label{GW0h2}
\Omega_{\mathrm{GW}, 0} h^2 = 0.83 \left(\frac{g_{*}}{10.75}\right)^{-1/3} \Omega_{\mathrm{r}, 0} h^2 \Omega_{\mathrm{GW}}\left(\eta_c, k\right),
\end{align}
where $\Omega_{\mathrm{r}, 0} h^2 = 4.2 \times 10^{-5}$ is the current radiation density, and the frequency $f$ is related to $k$ as
\begin{align}\label{fhz}
f = 1.546 \times 10^{-15} \frac{k}{1 \, \mathrm{Mpc}^{-1}} \, \mathrm{Hz}.
\end{align}
Through numerical calculations in Eq.~(\ref{GW0h2}) and utilizing Eq.~(\ref{fhz}) , we present the current SIGW energy spectrum in Figure~\ref{GW}. The frequency range of SIGW signals spans from \(10^{-4}\) Hz to \(5 \times 10^{-2}\) Hz, corresponding to \(\Omega_{\mathrm{GW},0} h^2\) values between \(10^{-12}\) and \(10^{-7}\). These signals fall within the sensitivity ranges of future GW detectors such as LISA, Taiji, and TianQin. This provides a promising avenue to probe PBHs and the early universe. 
In addition, based on the method proposed in~\cite{K. Inomata, S. Wang, Z. C. Chen, M. Sasaki and T. Suyama}, we compute the SGWB spectrum generated by binary PBH mergers, using the peak mass of PBH mass distribution and \( f_{\rm PBH}\) obtained in this work. The corresponding result is shown in Figure~\ref{SGWB}, covering a frequency range from \(10^6\) to \(10^{16}\,\mathrm{Hz}\) and energy density from \(10^{-18}\) to \(10^{-7}\). Since the estimated PBH mass in our model is relatively small, around \(10^{-12}\) \( M_{\odot} \) , the SGWB lies in the extremely high-frequency regime. Although current GW detectors are not sensitive to such high-frequency signals, future high-frequency GW observatories may be able to probe this class of SGWBs.

 \section{conclusions}
 \label{conclusion}

The formation of a significant abundance of PBHs requires the power spectrum of curvature perturbations to reach an amplitude of at least $\mathcal{O}(10^{-2})$. While observations of the CMB indicate that the power spectrum is nearly scale-invariant on large scales ($k \lesssim 1\,\mathrm{Mpc}^{-1}$), its amplitude on smaller scales ($k \gtrsim 1\,\mathrm{Mpc}^{-1}$) remains poorly constrained and may be substantially enhanced.
In this study, we investigate an inflationary model with non-minimal derivative coupling, in which the coupling function consists of a constant term and a periodic term. Numerical results show that the predicted tensor-to-scalar ratio and scalar spectral index are consistent with the latest P--ACT--LB--BK18 observational constraints. On certain small scales, the square of the sound speed exhibits periodic oscillations, transforming the curvature perturbation equation into a Mathieu equation within the horizon. The solutions to this equation demonstrate that curvature perturbations can grow exponentially under appropriate conditions, thereby confirming the effectiveness of the sound speed resonance mechanism in enhancing small-scale curvature perturbations.

When these amplified perturbations re-enter the Hubble horizon during the radiation-dominated era, regions with sufficiently high density contrast may collapse gravitationally, leading to the formation of PBHs. These PBHs could constitute a significant fraction of dark matter.
Moreover, PBH formation is accompanied by the generation of  SIGWs, whose energy density spectrum $\Omega_{\mathrm{GW}, 0} h^{2}$ exhibits characteristic multi-peak structures. Our results demonstrate that these GW signals fall within the sensitivity ranges of next-generation space-based detectors such as Taiji, TianQin, and LISA.
In addition, due to the abundance of PBHs, numerous black hole mergers are expected. Based on the peak mass of PBH mass distribution and \( f_{\rm PBH}\) derived in this work, we evaluate the SGWB resulting from these mergers. The SGWB exhibits a higher characteristic frequency than SIGWs and a single-peak energy density spectrum. Although current detectors are not sensitive to such high-frequency signals, future high-frequency GW observatories may provide a novel observational window into this regime.

It is worth noting that the formation of PBHs and the generation of SIGWs in this work have been analyzed under the assumption that curvature perturbations follow a Gaussian distribution. However, a more complete analysis should account for the effects of non-Gaussianity in curvature perturbations (see e.g., \cite{Chen2022,Cai2019,QingGuoHuang2013,Zhang2021,F.Zhang2022,fengge2020,Chul-Moon Yoo2019,G.Franciolini2018,Matthew2022,SamuelPassaglia2019,BravoRafael2018,Cai2018,VicenteAtal2018,VicenteAtal2019,P. Adshead,SPi,HVR,G.Domènech}).
In particular, in scenarios where the power spectrum is enhanced via parametric resonance, including non-Gaussian corrections is essential for reliably evaluating their impact on PBH abundance and the spectrum of SIGWs.  Thus, a more detailed investigation of the role of non-Gaussianity in the evolution of PBHs and SIGWs is therefore left for future work.

\begin{acknowledgments}
We sincerely appreciate Professor Puxun Wu of Hunan Normal University for his invaluable assistance with this work. Additionally, we acknowledge the support of the Chengdu Normal University Talent Introduction Scientific Research Special Project under Grant No. YJRC202443.
\end{acknowledgments}

\end{document}